# Collective sliding relaxor ferroelectricity across thousand-layer van der Waals superstructures

Nianze Shang[1]†, Cuiju Yu[2]†, Yucai Lin[1], Juan C. Arias-Muñoz[1], Jiajie Qi[3], Quanlin Guo[3], MD Gius Uddin[1], Youqiang Huang[1], Tuomas Pajunpää[1], Christopher P. Edwards[1], Rodolfo E. Canet-Albiach[1], Faisal Ahmed[1], Andreas C. Liapis[1], Yaoqiang Zhou[1], Zhibin Zhang[3], Hao Hong[3], Qing Dai[4], Andrea Alù[5], José L. Lado[2]*, Kaihui Liu[3]*, Zhipei Sun[1]*

[1] QTF Centre of Excellence, Department of Electronics and Nanoengineering, Aalto University, Espoo, Finland

[2] Department of Applied Physics, Aalto University, Espoo, Finland

[3] State Key Laboratory for Mesoscopic Physics, Frontiers Science Center for Nano-optoelectronics, School of Physics, Peking University, Beijing, China

[4] School of Materials Science and Engineering, Shanghai Jiao Tong University, Shanghai, China

[5] Photonics Initiative, Advanced Science Research Center, City University of New York, New York, NY, USA.

*Corresponding authors : zhipei.sun@aalto.fi, khliu@pku.edu.cn, jose.lado@aalto.fi

†These authors contributed equally to this work.

**Relaxor ferroelectrics are among the most important electromechanical materials owing to their exceptionally large electrostrictive responses. However, conventional relaxor behavior is generally associated with chemical complex and structural disorders, where local compositional fluctuations generate polar nanoregions, making the microscopic origin of it difficult to control and engineer. Here we report sliding relaxor ferroelectricity (S-RFE), a previously unknown relaxor state in which engineered stacking-phase heterogeneity enables cooperative interlayer sliding to generate collective relaxor behaviour. Through stacking-phase engineering in rhombohedral boron nitride, we create sliding domains that act as structural analogues of polar nanoregions and cooperatively align under an electric field through interlayer sliding. This collective state produces reversible electromechanical deformation of up to ~ −1.6% with negligible hysteresis in ~300-nm-thick crystals, corresponding to ~1000 atomic layers. The resulting electrostriction coefficient ($Q_{33}$ of ~ −1200 $m^4C^{-2}$) exceeds those of most crystalline electrostrictors by orders of magnitude. Unlike conventional relaxors, whose electromechanical response is often suppressed by dipolar freezing at low temperatures, S-RFE remains active down to 1.6 K and maintains giant electromechanical functionality. As proof of concept, we further demonstrate voltage-programmable optical modulation and computational spectrometry with nanometre-scale spectral resolution. Our results establish sliding relaxor ferroelectricity as a new route to relaxor behavior and reveal how atomic-scale interlayer motion can be amplified into collective mesoscale functionality across approximately one thousand layers, providing a materials-by-design platform for electromechanical, photonic, and quantum technologies based on van der Waals materials.**

## Introduction

Relaxor ferroelectrics (RFE) are a technologically crucial class of electromechanical materials characterized by nanoscale polar heterogeneity rather than long-range ferroelectric order[1]. Owing to their large electromechanical response, low hysteresis, and broad tunability, they have been widely explored for applications ranging from precise actuation and sensing to energy storage and conversion[2-6]. However, conventional relaxor behavior typically emerges from chemically complex and structurally disordered materials, where local compositional fluctuations generate polar nanoregions[7]. Such intrinsic crystallographic and chemical disorder makes the microscopic origin of RFE difficult to model, control, and engineer, while also posing challenges for material reproducibility, stability, miniaturization, and device integration[8, 9]. As a result, RFE has long been regarded as fundamentally linked to disorder. Whether relaxor behavior can emerge from an ordered and designable material system, rather than from chemical complexity, remains an unresolved question in ferroelectric and materials science.

Sliding ferroelectricity in van der Waals (vdW) materials has opened a fundamentally new paradigm for ferroelectric materials design. Unlike conventional ferroelectrics, which rely on ionic displacement to switch polarization, weak interlayer vdW interactions enable relative sliding between adjacent layers, allowing reversible switching of electric polarization through interlayer sliding[10]. Sliding ferroelectricity has been demonstrated in artificially stacked bilayer boron nitride, rhombohedrally stacked transition-metal dichalcogenides (TMDs), and a growing family of vdW materials, establishing a rich platform for exploring emergent ferroelectric phenomena at the atomic scale[11-19]. Remarkably, these systems represent the opposite limit of conventional relaxors: they are structurally simple, highly ordered, and microscopically controllable. This contrast raises a fundamental question of whether the highly ordered sliding ferroelectrics can support relaxor-like collective behavior without relying on chemical disorder.

To date, nearly all studies of sliding ferroelectricity have focused on atomically thin structures comprising only a few layers, where the phenomenon is primarily regarded as an interfacial effect. Although these studies have uncovered rich fundamental physics, it remains unknown whether interlayer sliding can become cooperative across hundreds or thousands of layers to generate emergent collective states and functionalities at mesoscopic length scales. In conventional RFEs, relaxor states provide a powerful mechanism for converting local polarization fluctuations into collective electromechanical responses[20]. Establishing an analogous collective state in sliding ferroelectrics would not only expand the concept of RFE beyond disordered materials but also bridge the gap between atomically thin interfacial ferroelectricity and bulk-scale electromechanical functionality. Whether cooperative interlayer sliding can give rise to such a collective relaxor state has remained unexplored.

Here we introduce and experimentally realize sliding relaxor ferroelectricity (S-RFE), a previously unexplored collective ferroelectric state in which relaxor behavior emerges from engineered stacking-phase heterogeneity and cooperative interlayer sliding, rather than from chemical disorder. Using rhombohedral boron nitride (rBN) as a model system, we create coexisting stacking phases that form sliding domains analogous to the polar nanoregions of conventional relaxor ferroelectrics. In contrast to conventional relaxors, where local polarization dynamics originate from compositional fluctuations, polarization switching in S-RFE is governed by interlayer sliding within these domains. Remarkably, cooperative sliding propagates across ~ 1000 atomic layers, transforming an interfacial sliding mechanism into an emergent collective state with mesoscale electromechanical functionality. This demonstrates that relaxor behavior can arise from structural design rather than crystallographic disorder, establishing a new route toward relaxor ferroelectricity. Leveraging this atomic-to-mesoscale transduction mechanism, we uncover giant electromechanical coupling in rBN and demonstrate its potential for programmable photonic applications, including optical modulation and computational spectral sensing.

Experimentally, we observe a thickness modulation in rBN that varies quadratically with the applied out-of-plane electric field, satisfying the defining characteristic of electrostriction, with a maximum reversible deformation of ~−1.6% at an electric field of ~200 MV/m. The corresponding electrostriction coefficient, $Q_{33} \sim -1200$ $m^4C^{-2}$, exceeds those of most crystalline electrostrictors by orders of magnitude and is supported by density functional theory calculations. Unlike conventional RFE, whose electromechanical response is frequently suppressed by dipolar freezing at low temperatures, S-RFE remains active down to 1.6 K, revealing an electromechanical mechanism that is intrinsically robust against cryogenic freezing. Furthermore, by integrating S-RFE rBN into a Fabry–Pérot cavity, we realize voltage-programmable optical modulation with a modulation depth of ~17%, enabling computational spectrometry with spectral resolution down to ~2 nm. As the first realization of sliding relaxor ferroelectricity, our work bridges atomically thin sliding ferroelectrics and collective mesoscale functionality, establishing a materials-by-design platform for electromechanical, photonic, and quantum technologies based on van der Waals materials.

## Emergence of a Sliding Relaxor Ferroelectric State

Conventional RFE derive their functionality from polar nanoregions embedded in a non-ferroelectric matrix, as illustrated in Fig. 1a. In S-RFE, we realize a structurally engineered relaxor architecture through stacking-phase heterogeneity in a van der Waals crystal. Rather than arising from chemical disorder, the relaxor state emerges from coexisting sliding ferroelectric domains with opposite polarizations that act as structural analogues of the polar nanoregions in conventional relaxors. These sliding domains provide the elementary building blocks of a collective relaxor state. As illustrated in Fig. 1b&c, S-RFE (using rBN as a model system) exhibits distinct stacking phases (ABC and CBA) as the sliding domains, with boundaries appearing only along the out-of-plane z direction. The polarization in each sliding domain originates from the out-of-plane dipole moment between vertically aligned boron and nitrogen atoms[21]. In Fig.1d, we show the top-view stacking structure of those two phases, along with other inactive phases possibly existing in the material, for comparison. When the external electric field is zero, the ABC and CBA phases are energetically degenerated (Fig. 1e). However, when a static electric field is applied along the −z (+z) direction, the oppositely oriented +z (−z) dipole of the CBA (ABC) stacking becomes destabilized. This lifts the energy degeneracy between the two polar stacking states and drives selective interlayer sliding, causing local sliding domains to align with the applied field. The collective alignment of many such domains forms the microscopic origin of the S-RFE state (Fig. 1f,g). Such selective sliding transitions are further confirmed by nudged elastic band (NEB) calculations (Fig. S1) and form the core mechanism underlying the electromechanical effect in rBN S-RFE crystals[22]. We conclude that S-RFE emerges from the interplay of three elements: coexisting polar sliding domains that act as structural analogues of polar nanoregions, field-driven interlayer sliding that couples stacking order to polarization, and cooperative domain evolution that converts local sliding events into a collective electromechanical response.

We successfully identify and observe the stacking-phase coexistence required to realize S-RFE from chemical vapor deposition (CVD) samples, based on the method developed in Ref. 23 (details in the SI). Due to the degenerate energy of ABC and CBA stacking, the reverse of the phases is triggered by the perturbative variation of growth parameters, leading to the formation of phase-coexisted S-RFE rather than the uniform ferroelectric structure.

Transmission electron microscopy (TEM) confirms this phase coexistence: the pink and blue parallelograms in Fig. 2a mark regions with opposite stacking phases corresponding to S-RFE-active domains. The thickness of each domain can vary from two layers (single AB bilayer) to several nanometers. Between these domains, there are also discontinuities consisting of mixtures of S-RFE inactive phases, such as AA', which is the dominating phase of hexagonal boron nitride (hBN). Fortunately, AA' can be readily distinguished from ABC↔CBA coexistence via second harmonic generation (SHG) measurements, as the in-plane B–N polarization remains aligned among different layers and the breaking inversion symmetry yields stronger SHG[23, 24]. As shown in Fig. 2c, rBN (primarily ABC and CBA phases) exhibits significantly stronger SHG than hBN (predominantly AA′ phase). Polarization-resolved SHG reveals sixfold symmetry in both parallel and perpendicular components, consistent with the intrinsic threefold in-plane symmetry of ABC and CBA phases (Fig. 2d). Out-of-plane polarization probed by piezo response force microscopy (PFM) further supports our conclusion: hBN shows no amplitude or phase response, indicating no intrinsic vertical polarization, whereas rBN displays an even-symmetric linear-like amplitude dependence on voltage and a phase reverse between ±90° when they are saturated above ±5 V (Fig. 2e,f). The PFM response exhibits two hallmark signatures of S-RFE. The absence of pronounced hysteresis rules out conventional ferroelectric switching (Fig. S2), while the even-symmetric field dependence is incompatible with linear piezoelectricity. Instead, both observations are consistent with relaxor-like electrostriction arising from cooperative sliding domains, where the strain scales quadratically with the applied electric field.[25] (detailed analysis for the PFM data can be found in SI, Supplemental Text).

The mechanism accounting for the electromechanical responses can be further confirmed by theoretical simulations based on DFT using the S-RFE model (details in SI, Materials and Methods). As discussed in Fig. 2g, under an external electric field, the energy degeneracy between the ABC and CBA phases is lifted, and the latter becomes energetically favored. This energy difference drives the sliding process from ABC to CBA. The electric field induced ferroelectric alignment further drives the deformation of the rBN. Specifically, the impact of the ferroelectric order on the structural rBN deformation can be inferred directly from first principles methods. For this purpose, we consider a minimal CBA(ABC) trilayer model under the +z(-z) electric field and calculate its thickness variation ($\Delta t = (t - t_0)/t_0$, where $t_0$ and t denote the effective thickness before and after applying the electric field, respectively) as the electric field strength is increased. We obtain that $\Delta t$ scales quadratically with the net out-of-plane polarization ($P_z$): $\Delta t \propto P_z^2$ (Fig. 2h)[2]. By combining the first-principles calculations, we derive the $\Delta t$–$P_z$ relationship in rBN S-RFE, which yields a symmetric quadratic curve and agrees well with the behavior of relaxor ferroelectrics. Phonon spectra in rBN are also stable in the presence of the applied field, implying a good robustness of the model (see Materials and Methods for detailed information).

Together, TEM, SHG, PFM, and first-principles calculations establish a microscopic picture of S-RFE in which structurally engineered sliding domains replace the polar nanoregions of conventional relaxors. The field-driven cooperative evolution of these domains generates a collective electromechanical state that is fundamentally distinct from both conventional relaxor ferroelectrics and previously reported atomically thin sliding ferroelectrics.

## From Collective Sliding to Giant Electrostriction

A defining consequence of S-RFE is that the collective evolution of sliding domains should convert local interlayer displacement into a macroscopic electromechanical response. To test this prediction, we quantify the field-induced thickness modulation of rBN using optical reflection spectroscopy (schematic in Fig. 3a). rBN flakes are integrated into Fabry–Pérot (F–P) cavity devices made of Au/rBN/Au sandwiched heterostructures via lithography and dry-transfer processes (details in Methods). The Au layers serve as both electrical contacts and top/bottom cavity mirrors. Under a perpendicular electric field, an increase or decrease in cavity thickness produces a corresponding red or blue shift in the F–P resonant peaks, directly revealing thickness tunability (see inset of Fig. 3a). Fig. 3b shows the fabricated rBN S-RFE device 1, where the Au/rBN/Au cavity area exhibits a distinct green color contrast. AFM measures the rBN thickness as ~260 nm (Fig. S3), matching the second-order F–P resonant mode for rBN with estimated refractive index $n \approx 2$, consistent with reported values[26]. This is confirmed by the cavity's reflection spectrum, showing a peak at ~564 nm and valleys at ~480 nm and ~650 nm (Fig. 3c). Transfer-matrix simulations for the Au/rBN/Au structure on a 300 nm $SiO_2$/Si substrate reproduce the experimental peak and valley positions, with intensity deviations in the violet and infrared spectral regions attributed to optical component dispersion.

Reflection spectra of the fabricated Au/rBN/Au cavity under different perpendicular DC voltages are collected and mapped (Fig. 3d). The applied voltage ($V$) is defined as $V_{top} - V_{bottom}$ (where $V_{top}$ and $V_{bottom}$ denote the voltages of the top and bottom contacts, respectively), and the displayed data is smoothed to remove high-frequency noises for a better determination of the peak position. Dashed contour lines at reflection intensities ~0.63, 0.72, and 0.9 are added to reveal curvatures for contour lines between 500–800 nm. In Fig. 3e, we show the spectra extracted for 0 V, −50 V and +50 V (curves under more voltages in Fig. S4). The observation that both positive and negative electric fields produce the same sign of deformation immediately identifies the response as electrostrictive rather than piezoelectric. By contrast, the hBN-based reference device shows no voltage dependence (Fig. S5), excluding Maxwell stress effects and confirming that the deformation arises from the S-RFE in rBN[27]. A slight asymmetry between positive and negative bias is observed in Fig. 3f. The origin of the electrostriction can be well explained by the unique ABC-CBA phase coexistence and interfacial sliding mechanism of S-RFE, as discussed in Fig. 2g and h, and the asymmetry can be attributed to the unequal ratio between ABC and CBA phases in the original rBN crystal. Remarkably, this atomic-scale interlayer sliding mechanism produces a macroscopic electromechanical response across a ~300-nm-thick crystal, corresponding to approximately one thousand atomic layers. Such behavior demonstrates that cooperative sliding is not confined to individual interfaces but can propagate collectively through thick van der Waals crystals

This S-RFE induced electromechanical response becomes more pronounced when the blue shifts are converted to the thickness reduction $\Delta t$ via $\Delta t = (\Delta\lambda_m/\lambda_m)\cdot t$, where $\Delta\lambda_m$ is the wavelength shift of the $m^{\text{th}}$ order resonant peak, and $t$ is the thickness of the rBN. In our case, $m$ = 2 for the peak at ~564 nm (Fig. 3e). This thickness reduction behavior is consistent with the S-RFE mechanism, in which local sliding domains align under either field polarity and collectively evolve toward a lower-energy state, generating a macroscopic deformation. At −50 V, the maximum thickness change is ~ −4.2 nm, corresponding to a deformation ratio (or strain) $s_{zz}$ ~ −1.63%, larger than most of the reported electrostrictors[27, 28] (Fig. 3f). Note that the observation of ~1.6% reversible strain demonstrates that the cooperative sliding across approximately one thousand atomic layers can amplify atomic-scale interfacial motion into a giant collective electromechanical response. The equivalent electrostriction coefficient $Q_{33}$ is then obtained by fitting the $\Delta t$–$V$ curve to a quadratic function, as defined by strain per unit of electric polarization intensity ($s_{zz} = Q_{33}{P_z}^2$, Voigt notation 3 ≡ zz)[29]. Including the out-of-plane dielectric constant $\varepsilon_\perp$ into the electrostriction metrics[30], we derive the polarization-based electrostriction coefficient by the equation $s_{zz} = Q_{33}\varepsilon_0^2(\varepsilon_\perp - 1)^2{E_z}^2$. The extracted electrostriction coefficient reaches $Q_{33} \sim -1200$ m$^4$C$^{-2}$ (detailed calculations and discussion in SI), one order of magnitude larger than most reported giant crystalline electrostrictors, and qualified as 'giant electrostriction' according to the commonly recognized equation $|Q| > 2.37(\Theta/\varepsilon_0\varepsilon_r)^{0.59}$, marking the performance of the material surpassing the limit set by elasticity and electric permittivity (where Θ is the elastic compliance)[27]. Such an extraordinary electrostrictive response reveals that S-RFE is a highly cooperative phenomenon extending throughout the entire ~1000-layer crystal, rather than a localized interfacial effect. The resulting collective electromechanical state represents a new regime of deformation engineering in van der Waals materials, where nanoscale interlayer motion is amplified into a giant macroscopic response. Comparisons of the deformation ratio and $|Q_{33}|$ with high-performance electrostrictors, e.g., ceria, $Bi_2O_3$ and LAMOX reported so far[31-39], are shown in Fig. 3g. With both order-of-magnitude larger $|Q_{33}|$ and deformation, S-REF demonstrates a strong potential as a platform for electromechanical applications.

The quadratic deformation–voltage trend matches well with first principles calculations (Fig. 2h), supporting an S-RFE origin for the electromechanical response. Log–log plots give a power-law dependence of $P^{1.7}$ experimentally and $P^{1.9}$ in calculations (Fig. S6). Furthermore, the $Q_{33}$ obtained by calculation is ~ $-2200$ m$^4$C$^{-2}$ by fitting the equation $\Delta t = Q_{33}P_z^2$, closely matching the experiment value (considering the $|Q_{33}|$ obtained from different devices ranges between 1000-7000 m$^4$C$^{-2}$ because of the variation of crystal quality) and supporting our optical characterization.

## Robust Cryogenic S-RFE

The temperature-dependent measurements reveal the remarkable robustness of the S-RFE response under cryogenic conditions. As shown in Fig. 4 a&b, at 1.6 K, the F-P resonances in reflectivity spectra (from Device 2, resonance peaks centered at ~524 nm and 775.5 nm) blueshift symmetrically with applied positive or negative voltage, indicating reversible thickness modulation. The extracted deformation follows a clear quadratic dependence on the applied electric field. To examine the temperature evolution of the electromechanical response, deformation-voltage characteristics are measured from 1.6 K to 300 K (Fig 4c). Two representative temperature points: 1.6 K and 80 K, are selected and presented with original deformation-voltage data points, along with the quadratic fitting curves (Fig. 4d,e, details of other temperature points are in Fig. S7). It is noted that a different voltage range is applied at different temperatures during the measurement, keeping the maximum deformation achieved roughly the same. The quadratic terms of each fitted curve are then extracted and plotted versus temperature, which represents the deformation under per voltage squared (Fig. 4f). Meanwhile, we calculate the $Q_{33}$ by assuming the permittivity of rBN stays nearly constant across the whole temperature regime. As a function of temperature, the absolute value of $Q_{33}$ gradually increases with temperature, indicating enhanced electromechanical response at elevated temperatures. This trend likely arises from increased thermal activation that facilitates interlayer sliding while preserving the same underlying electromechanical mechanism. Based on a qualitative picture that the deformation is proportional to the number of sliding-possible interfaces, which is affected by thermal activation, we fit the temperature-dependent $Q_{33}$ with an exponential curve which matches well to experiment data across the whole temperature range (Details of exponential fitting and discussion are in SI).

Importantly, giant electrostriction persists down to 1.6 K, where $Q_{33}$ remains on the order of $-1000$ $m^4C^{-2}$. Unlike conventional relaxor ferroelectrics[40], whose electromechanical response is suppressed as dipolar dynamics freeze at low temperature, S-RFE retains nearly the same order of electromechanical coupling throughout the entire temperature range investigated. The persistence of giant electrostriction at cryogenic temperatures provides compelling evidence that S-RFE constitutes a distinct class of relaxor ferroelectricity, in which collective interlayer sliding rather than dipolar reorientation serves as the primary order parameter[41, 42]. This unique temperature robustness opens opportunities for large-strain actuation and strain engineering in cryogenic photonic, superconducting, and quantum-information platforms.

## Applications of S-RFE

The combination of giant electrostriction, wide optical transparency arising from the large bandgap of rBN [43] (~5.9 eV), and electrically tunable cavity resonances enables S-RFE to function as a new platform for active photonics. As a proof-of-concept, we demonstrate computational spectral detection based on voltage-controlled optical modulation (Fig. 5a). S-RFE device 3 was fabricated with rBN thickness of ~320 nm, showing similar resonant peak shifts and thickness tuning under an applied electric field (Fig. S8). However, the voltage corresponding to the maximum thickness shifts from +5 V (device 1) to approximately –10 V, likely due to different ABC/CBA phase ratios during the crystal synthesis. Different wavelengths generate distinct voltage-dependent modulation profiles because the electrically induced cavity resonance shift modifies the reflectivity spectrum in a wavelength-selective behavior. In experiments, monochromatic light (500–600 nm) is focused onto the cavity device, and the reflected intensity is measured by a photodetector (Detailed setup in Fig. S9). The device exhibits a clear response that varies with wavelength (Fig. 5b), with, for example, the reflected light intensity at 540 nm and 580 nm signals showing opposite voltage-dependent trends and markedly different modulation due to the opposite reflectivity changes induced by spectral blue shift. The modulation depth, defined as $(I_{max} - I_{min})/I_{max}$, where $I_{max}$ and $I_{min}$ stand for the modulated light intensity measured at maximum or zero voltages, reaches ~ $-17\%$ (520 nm) and ~14% (560 nm) at 40 V. The results for different wavelengths are presented in Fig. 5c (blue dots and line), highlighting its good capability for wavelength detection. Importantly, the measured modulation depth is not limited by the intrinsic S-RFE response, but by the relatively low cavity quality factor (Q ≈ 8) arising from mirror roughness and optical absorption. Numerical simulations indicate that nearly 100% modulation depth becomes achievable when Q exceeds 50 (Fig. S10f).

The high-contrast, wavelength-selective modulation with strong voltage dependence of our S-RFE rBN device enables accurate spectral reconstruction[44-46]. Therefore, the wavelength-dependent modulation profiles constitute a unique optical fingerprint for each input spectrum, enabling spectral information to be encoded into a sequence of voltage-dependent intensity measurements. By systematically tuning the voltage, the reflected light intensity encodes wavelength information. To experimentally realize this operation, we first collect a full set of responsivity matrices (reflected light intensity versus wavelength and voltage). Fig. 5d shows the 2D colormap of the responsivity matrix, revealing strong voltage dependence and wavelength variability, further highlighted in the first-derivative map (Fig. 5e). Second, a neural network is trained on the voltage-tunable intensity profiles from known spectra with both monochromatic (500–600 nm) and broadband with multi-peaks. Finally, voltage-dependent reflected light intensities from unknown inputs are used for spectra reconstruction. As shown in Fig. 5f, the reconstructed peaks (dashed lines) closely match the input reference spectrum measured from a commercial spectrometer. The correlation between reconstructed and actual wavelengths is excellent ($R^2 \approx 0.99$, Fig. 5g). Multiple-peak spectral discrimination is also achieved, with a best spectral resolution of ~2.5 nm for two closely spaced peaks (Fig. 5h). These results establish S-RFE as a new class of active photonic materials, in which giant electrically driven mechanical deformation is directly converted into optical functionality. Beyond computational spectroscopy, the large, reversible, and low-power cavity tuning enabled by S-RFE opens opportunities for tunable optical filters, reconfigurable photonics, programmable metasurfaces, and integrated photonic systems.

## Conclusion

In conclusion, we have established sliding relaxor ferroelectricity as a distinct class of relaxor ferroelectricity. In S-RFE, structurally engineered sliding nanoregions replace the polar nanoregions of conventional relaxor ferroelectrics, and collective interlayer sliding serves as the primary order parameter, replacing ionic displacement and dipolar reorientation that underpin conventional ferroelectric and relaxor responses. Through stacking-phase engineering in rhombohedral boron nitride, we demonstrate that cooperative sliding dynamics across approximately one thousand atomic layers give rise to a collective electromechanical response characterized by reversible deformation of up to ~−1.6%, an effective electrostriction coefficient of $Q_{33} \sim -1200\ m^4C^{-2}$, and robust operation from room temperature down to 1.6 K. Beyond establishing a new route to RFE based on structural design rather than chemical disorder, our work bridges atomic-scale interlayer dynamics and collective device functionality, establishing structural phase engineering as a powerful materials-by-design strategy for creating emergent functional states in vdW solids. The concept of S-RFE opens new opportunities for slidetronics, while providing a general framework for translating interfacial sliding into emergent electromechanical, optical, and quantum functionalities across multiple length scales.

## Acknowledgements

We thank V. Asadchy, T. Hasan, Z.S. Chen for the valuable discussions. We acknowledge the provision of facilities and technical support from the Otaniemi research infrastructure (OtaNano-Micronova Nanofabrication Centre and OtaNano-Nanomicroscopy Centre).

**Funding**: This work was supported by the the BARI program, the Department of Defense and the Research Council of Finland (353364, 360411, 359009, 365686, and 367808), the Research Council of Finland Flagship Programme (320167, PREIN), InstituteQ, the Finnish Quantum Flagship, ERC (834742, 101170477), the National Natural Science Foundation of China (92577201, 52025023) and the EU Marie Skłodowska-Curie Actions (ALTOS). This work is part of the Finnish Centre of Excellence in Quantum Materials (QMAT).

**Author contributions**: Z.S. conceived of the ideas during discussion with N.S. N.S. designed the experiments, fabricated rBN devices and carried out the characterization and measurements. C.Y. carried out DFT calculations. Y.L. developed the reconstruction codes for the computational spectrometer. J.Q. synthesized the rBN crystal. J.C.A-M., Y.H. commented on the experimental results and helped with data analysis. J.C.A-M helped with the graphic design. Q.G performed TEM characterization. G.U., T.P., A.C.L. helped to fabricate the rBN S-RFE devices. T.P., C.E. carried out transfer matrix calculations. N.S. and C.Y. wrote the manuscript. H.H., A.A., Q.D., R. C-A. provided insightful comments on the manuscript writing. Z.S., K.L., J.L. supervised the research. All authors participated in the scientific discussion extensively and contributed to the writing of the manuscript.

N.S. and C.Y. contributed to this work equally.

**Competing interests:** The authors declare that they have no competing interests.

**Data and materials availability:** All data are available in the main text or the supplementary materials.

**Supplementary Materials**

Materials and Methods

Supplementary Text

Figs. S1 to S13

Tables S1

References (47-55)

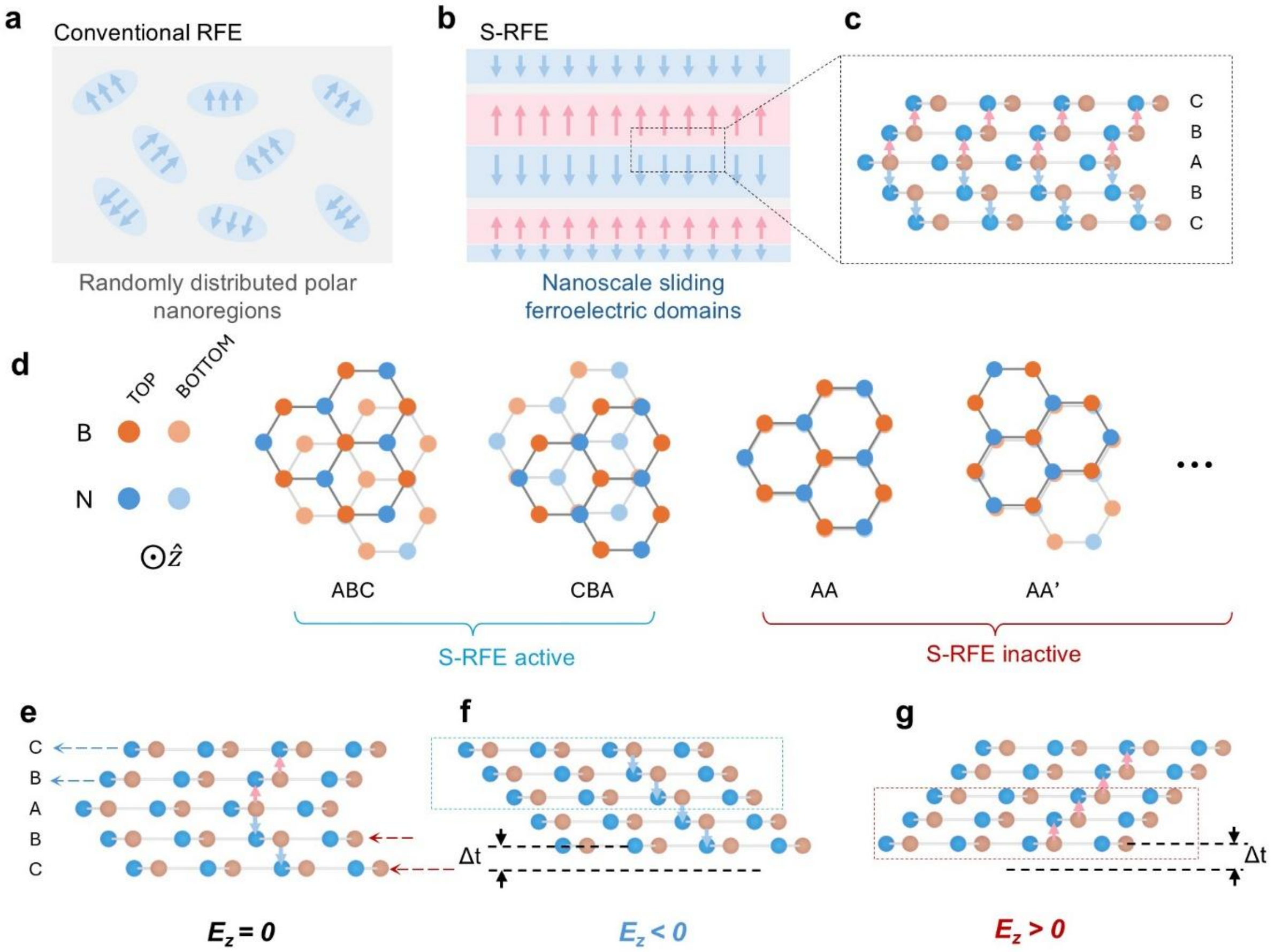

**Figure 1. Concept of sliding relaxor ferroelectricity (S-RFE).** **a.** Schematic of conventional relaxor ferroelectricity (RFE), in which polar nanoregions are randomly distributed within a non-polar matrix. **b.** Schematic of sliding relaxor ferroelectricity (S-RFE), where stacking domains with opposite out-of-plane polarizations coexist within a van der Waals crystal, forming sliding nanoregions. **c.** Atomic structure at the interface between ABC and CBA stacking domains. The two phases possess opposite out-of-plane polarizations and form the fundamental active unit of S-RFE. **d.** Illustration of different stacking phases: ABC, CBA, AA and AA'. All ABC and CBA phases are S-RFE active because of the net polarization they generate and the sliding enabled transition between each other. AA, AA', and other possible phases are S-RFE inactive. **e,f,g**. The mechanism for selective interlayer sliding in S-RFE rBN illustrated by a five-layer model. **e.** The original CBABC stacks when external out-of-plane electric field $E_z = 0$. **f.** $E_z < 0$ case, CBA phase slides to ABC phase (indicated by blue dashed arrows in e), the whole stack adopts a unified +z polarization, and a thickness reduction due to the interlayer distance decreasing. **g.** $E_z > 0$ case, ABC phase slides to CBA phase (indicated by red dashed arrows in e), the whole stack adopts a unified -z polarization, and also a thickness reduction. The collective evolution of these sliding domains gives rise to a macroscopic electromechanical response.

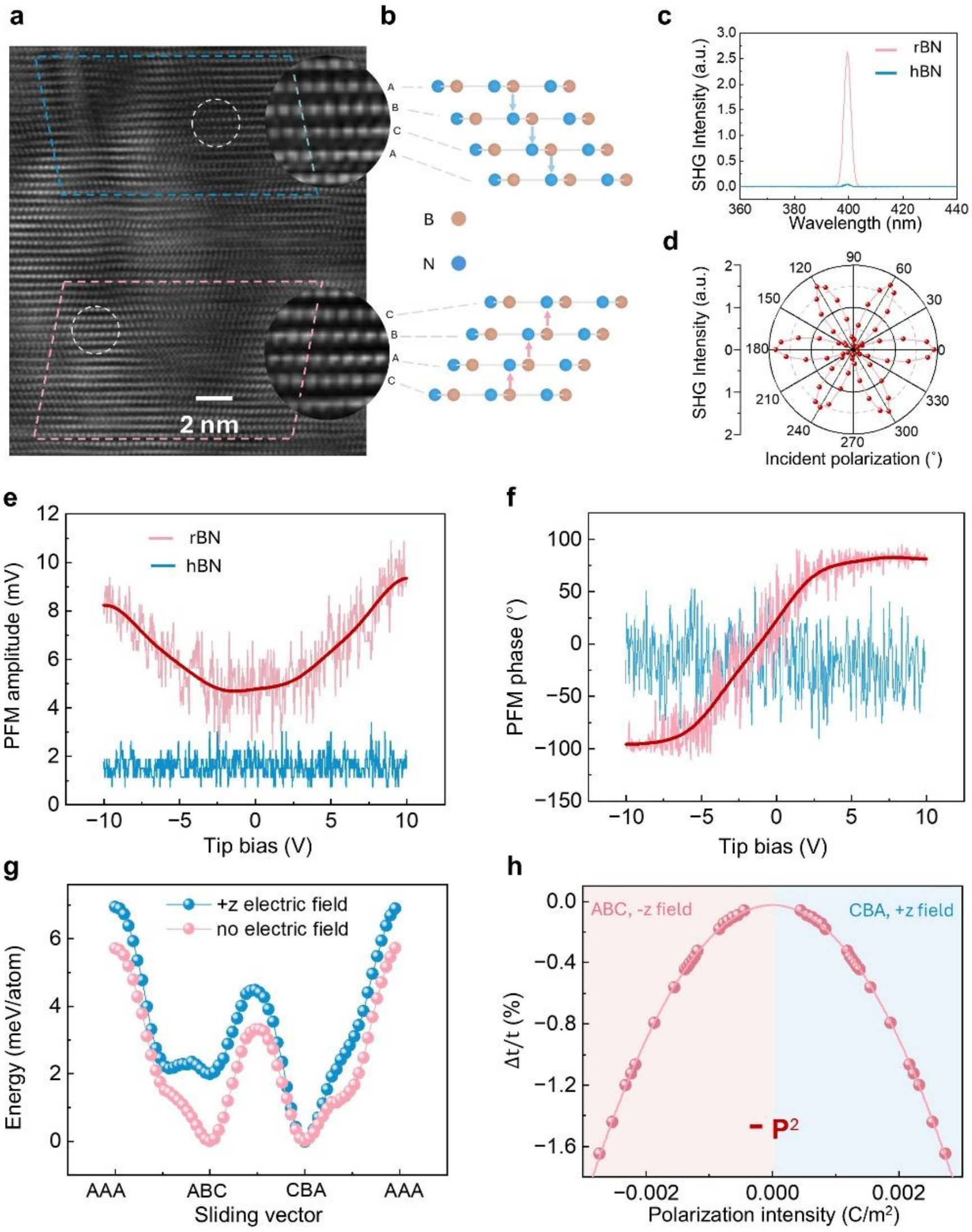

**Figure 2. Experimental observation of S-RFE in rBN**. **a.** Direct observation of coexisting ABC (blue dashed box) and CBA (red dashed box) stacking domains with TEM. **b.** Schematics of the ABC (top) and CBA (bottom) phases corresponding to the TEM observation. **c.** SHG confirmation of broken inversion symmetry in rBN. **d.** Excitation polarization dependent SHG intensity of rBN. The sixfold paddle indicates the threefold in-plane symmetry of ABC/CBA phases of rBN. **e.f.** Confirmed out-of-plane polarization with **e.** amplitude and **f.** phase signals of PFM in rBN. **g.** Calculated energy per atom vs different sliding vectors with/without an external electric field. DFT reveals a bistable sliding pathway between ABC and CBA phases. The CBA stacking becomes energetically favored in the presence of a positive vertical electric field. **h.** Calculated thickness variation of rBN with phase coexistence under an external electric field. Under the +z(-z) field, ABC-CBA coexisting phases transform to a unified CBA (ABC) phase by interfacial sliding. The post-sliding rBN undergoes a quadratic thickness variation as the electric field intensity increases.

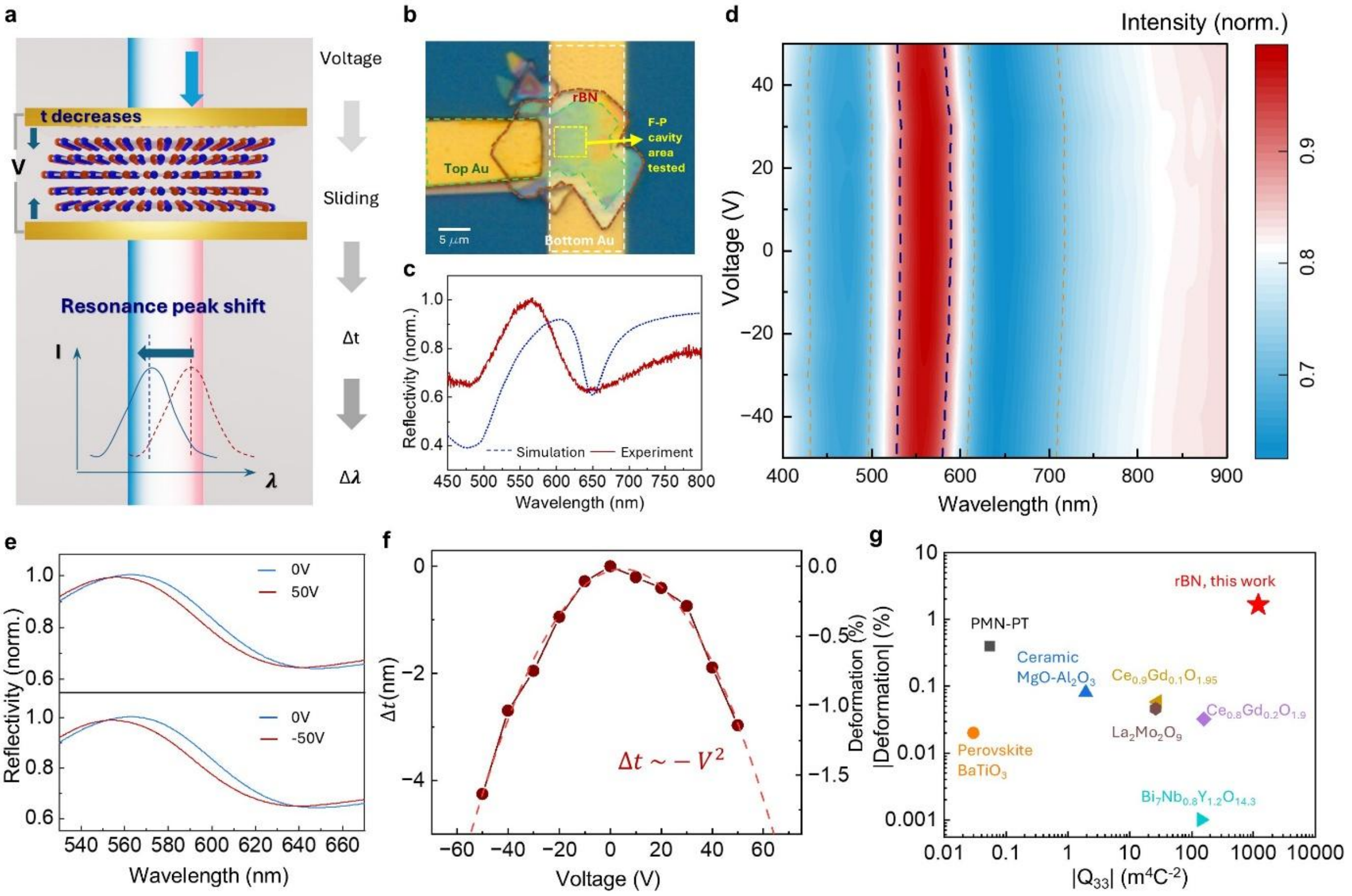

**Figure 3. Giant electromechanical response of S-RFE rBN**. **a.** Schematics of the experiment. White light is focused on the S-RFE device with top and bottom Au contacts. The transmitted (or reflected) light spectra are then measured and analyzed. The red (or blue) shift of the spectra corresponds to the increase (decrease) of the thickness of the F-P cavity induced by voltage. **b.** Image of S-RFE device 1. The green color of the device indicates that the peak position of the F-P cavity reflectivity spectrum is located around 500-600nm. The rectangular box marks the F-P cavity area measured and tested. **c.** Reflection spectra of S-RFE device 1 (red, solid), compared to the calculated spectra by the transfer matrix method (blue, dashed). **d.** 2d colormap of reflectivity spectra of S-RFE Device 1 at different voltages. The sign of the voltage is defined as $V_{top}$-$V_{bottom}$. Contour lines for values 0.72 (orange) and 0.9 (black) are marked in the map, with curvatures indicating the device's tunability under external voltage. **e.** Reflectivity spectra curves of rBN F-P Device 1 under -50V, 0V and 50V. Both directions of voltage induce blue shifts of the reflection spectra curve, attributed to the reduction of device thickness. **f.** Calculated thickness change of the rBN flake under external voltage (points), which can be fitted by a quadratic curve (dashed line). **g.** Comparison of the maximum achievable deformation ratio (absolute value) and calculated electrostriction coefficient $|Q_{33}|$ of rBN with previously reported conventional and giant electrostriction materials.

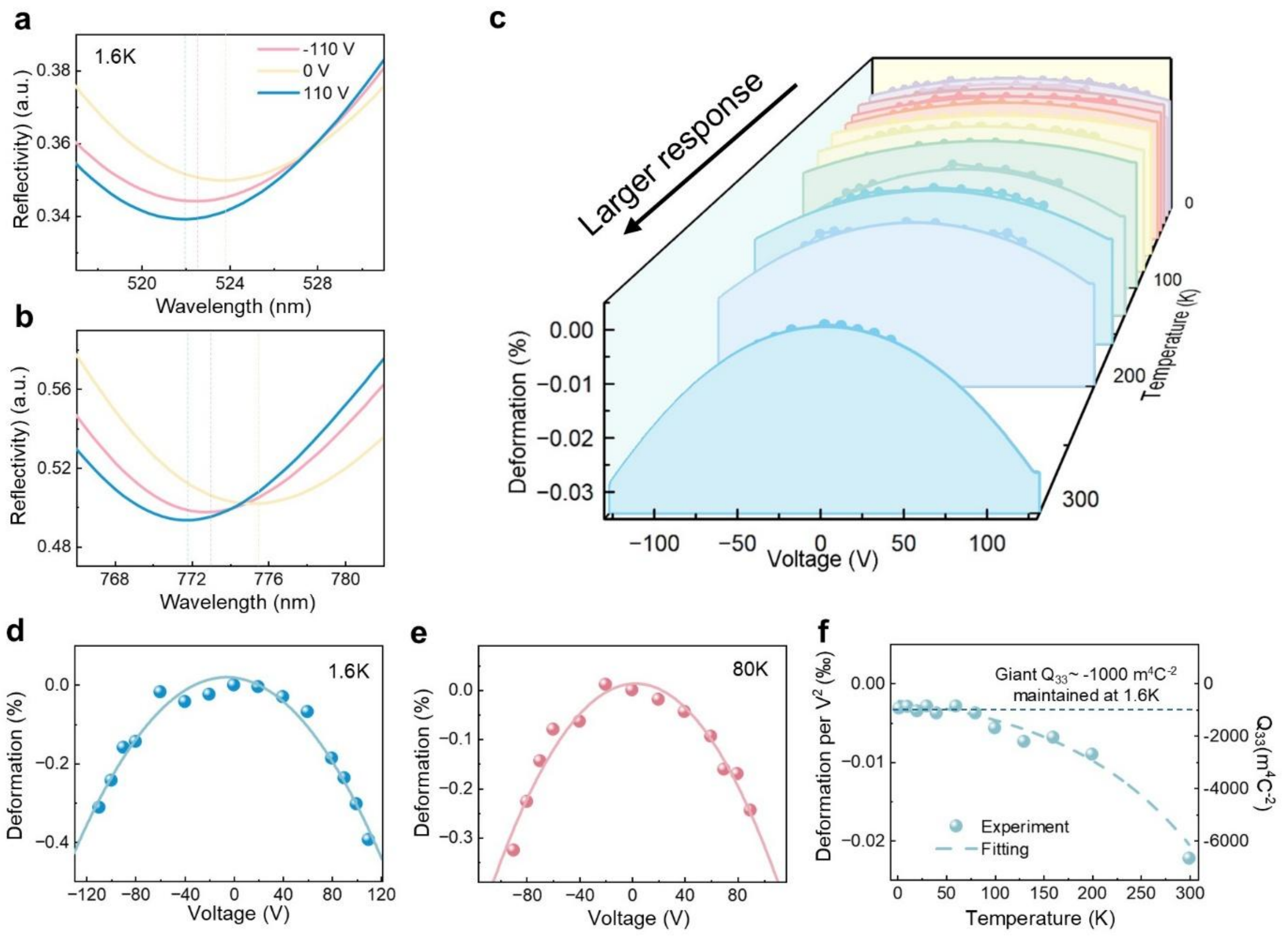

**Figure 4. Robust cryogenic S-RFE**. **a, b.** Voltage-dependent reflectivity spectra of S-RFE Device 2 measured at 1.6 K. Fabry-Pérot resonances centered at ~524 nm (a) and ~775.5 nm (b) exhibit symmetric blueshifts under positive and negative bias, indicating reversible electrically driven thickness modulation. **c**. Deformation-voltage characteristics measured from 1.6 K to 300 K. Solid lines represent quadratic fits to the experimental data. The electrostrictive response persists across the entire temperature range investigated. **d,e.** Representative deformation-voltage curves at 1.6 K (d) and 80 K (e). Symbols denote experimental data and solid lines are quadratic fits, confirming the electrostrictive nature of the response. Different voltage ranges were used at different temperatures while maintaining a comparable maximum deformation. **f**. Temperature dependence of the quadratic deformation coefficient (left axis) and the corresponding electrostriction coefficient $Q_{33}$ (right axis), calculated assuming a temperature-independent dielectric permittivity. The dashed line is an exponential fit. Although the electrostrictive response gradually increases with temperature, a giant electrostriction coefficient on the order of $10^3$ $m^4C^{-2}$ is maintained down to 1.6 K, demonstrating the exceptional cryogenic robustness of sliding relaxor ferroelectricity.

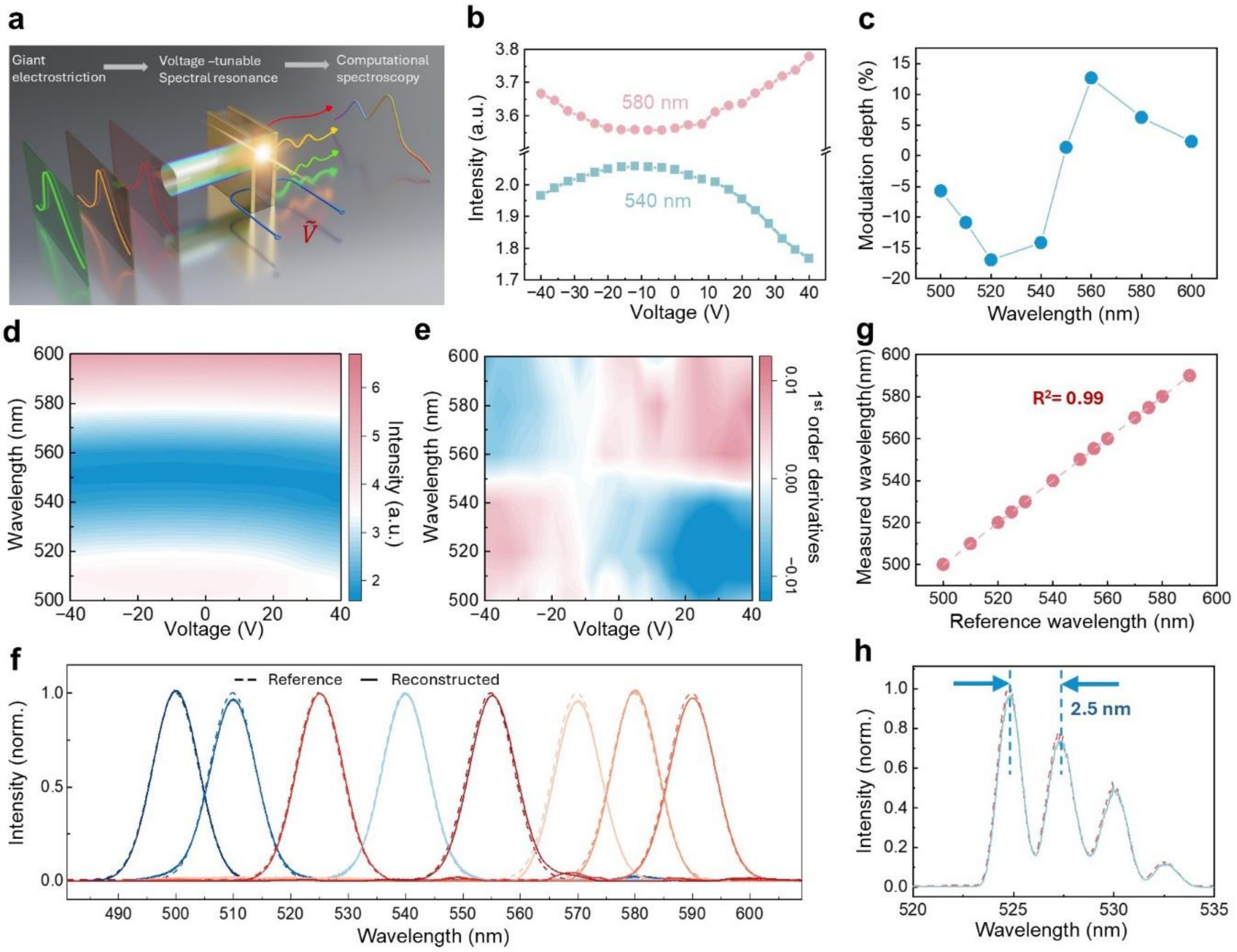

**Figure 5. Photonic applications of S-RFE. a.** Concept of an S-RFE based computational spectrometer. Voltage controlled electrostriction shifts the cavity resonance and generates wavelength-dependent optical responses that serve as spectral fingerprints. **b.** Modulation curves at two representative wavelengths of 540 and 580 nm, showing distinct trends and intensities. **c.** Extracted modulation depth for various wavelengths from 500 to 600 nm, with a maximum of ~ −17% (520 nm) and ~14% (560 nm) achieved at 40 V. **d.** 2D colormap of the intensity of reflected light dependent on wavelength and voltage, used as the responsivity matrix of the reconstructive algorithm. **e.** $1^{st}$ order derivatives of d. Rather distinct features in different sectors indicate good performance when used in the reconstruction process. **f.** Comparisons of reconstructed spectra (solid lines) and reference spectra (dashed lines) measured by a commercial spectrometer. **g.** Peak positions by reconstruction versus the real input peak position, showing excellent reconstruction fidelity. **h.** Multiple-peak spectrum measurement result (solid line) in contrast to its input reference spectrum (dashed line), demonstrating a spectral resolution of 2.5 nm.